\documentclass[runningheads]{llncs}

\usepackage[T1]{fontenc}
\usepackage{amsmath}
\usepackage{graphicx}
\usepackage{tabularx}
\usepackage{multirow}
\usepackage{multicol}
\usepackage{hyperref}
\usepackage{url}

\begin{document}
	
	\title{Activity Coefficient-based Channel Selection for Electroencephalogram: A Task-Independent Approach}
	\titlerunning{Activity Coefficient-based Channel Selection}
	
	\author{Kartik Pandey\orcidID{0009-0001-4933-6598} \and
		Arun Balasubramanian\orcidID{0009-0006-7910-6458} \and
		Debasis Samanta\orcidID{0000-0002-6104-3771}}
	\authorrunning{K. Pandey et al.}
	
	\institute{Indian Institute of Technology, Kharagpur, West Bengal, India \email{kartik4323@kgpian.iitkgp.ac.in}\\
	}
	
	\maketitle
	
	\sloppy
	
	\begin{abstract}
		Electroencephalogram (EEG) signals have gained widespread adoption in brain–computer interface (BCI) applications due to their non-invasive, low-cost, and relatively simple acquisition process. The demand for higher spatial resolution, particularly in clinical settings, has led to the development of high-density electrode arrays. However, increasing the number of channels introduces challenges such as cross-channel interference and computational overhead. To address these issues, modern BCI systems often employ channel selection algorithms. Existing methods, however, are typically task-specific and require re-optimization for each new application. This work proposes a task-agnostic channel selection method, Activity Coefficient-based Channel Selection (ACCS), which uses a novel metric called the Channel Activity Coefficient (CAC) to quantify channel utility based on activity levels. By selecting the top 16 channels ranked by CAC, ACCS achieves up to 34.97\% improvement in multi-class classification accuracy. Unlike traditional approaches, ACCS identifies a reusable set of informative channels independent of the downstream task or model, making it highly adaptable for diverse EEG-based applications.
		
		\keywords{Neurotechnology \and Deep Learning \and Brain–Computer Interface \and Channel Selection \and EEG Signal Processing}
	\end{abstract}
	
	\section{Introduction}
	\label{sec:1}
	
	The integration of high-density electrode arrays into Electroencephalogram (EEG) caps \cite{bell2012using} has significantly enhanced spatial resolution, enabling more precise localization of neural activity and advancing our understanding of brain function. However, this improvement also brings challenges—such as increased cross-channel interference and heightened risk of overfitting due to larger data volumes \cite{Baig2020}. These challenges are particularly critical in brain–computer interface (BCI) applications, the focus of this work. Unlike clinical EEG systems, where dense configurations support diagnostic precision \cite{Stoyell2021}, BCIs demand lightweight, real-time, and task-adaptive pipelines. Accordingly, the proposed channel selection method targets BCI contexts and is not designed for clinical diagnostics.
	
	To address the computational and generalization limitations introduced by high-resolution EEG data, it is common practice to select a representative subset of channels for downstream processing \cite{abdullah2022eeg}. Principal component analysis (PCA) \cite{1615389} is one such method, which splits the channels into its principal components and selecting them based on the importance. Most modern channel selection algorithms are task-specific, relying on classifier performance to iteratively identify and retain informative channels \cite{alotaiby2015review}. More recent examples of such data-driven methods include, Mutual Information (MI) \cite{wang2019channel}, Cross-Correlation Discriminative Criterion (XCDC) \cite{yu2021cross}, and Sequential Backward Floating Search (SBFS) \cite{tang2022eeg}. Deep learning-based approaches like Efficient Channel Attention Network (ECA-Net) \cite{wang2020eca} and multi-objective methods like non-dominated sorting genetic algorithm II (NSGA-II) \cite{moctezuma2020eeg} have advanced the field with greater improvements in performance, yet still depend on task-specific performance metrics.
	
	This study introduces a task-independent(TI) strategy, Activity Coefficient-based Channel Selection (ACCS), which eliminates reliance on downstream classifiers or labels by utilizing the activity level of the channels. ACCS is guided by four key goals:
	
	\begin{itemize}
		\item Define a metric to quantify the activity level of each channel.
		\item Design an algorithm to discard redundant channels based on data characteristics alone.
		\item Demonstrate dimensionality reduction with preserved or improved classification performance.
		\item Compare ACCS against existing task-dependent methods.
	\end{itemize}
	
	The core idea is to identify active channels whose signal levels exceed the resting-state baseline. To this end, we propose a new metric, the Channel Activity Coefficient (CAC), derived from the performance of a CNN model trained to distinguish between resting and active EEG states. Channels with high CAC values are selected and validated through multi-class classification using existing models. Improved classifier performance with fewer channels underscores the effectiveness of the proposed method.
	
	\section{Proposed Methodology}
	\label{sec:2}
	
	\begin{figure}[!tb]
		\centering
		\includegraphics[width=0.8\textwidth]{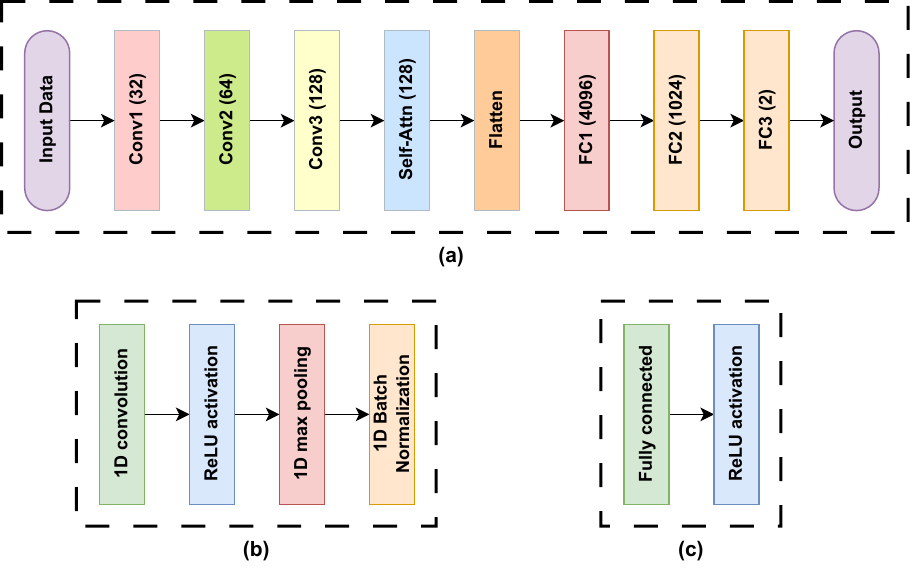}
		\caption{Overview of the proposed CNN: (a) main pipeline with output dimensions in brackets; (b) structure of each CNN block; (c) structure of the first two FC blocks. The self-attention layer uses a softmax activation.}
		\label{fig:1}
	\end{figure}
	
	The proposed pipeline first preprocesses signals to remove noise and artifacts, then feeds them to a CNN architecture (Fig.~\ref{fig:1}) trained to discriminate between active and resting state EEG signals. The CNN comprises three 1D convolutional blocks, each containing convolution (kernel size 3, stride 1, padding 1), ReLU activation, max pooling for dimension reduction, and batch normalization for stable feature distribution and faster training. Output channels for each block are shown in the figure.
	The convolutional blocks are followed by a single-head self-attention layer with linear projections for query, key, and value (hidden dimension 128) that operates over the temporal dimension to focus on informative time steps. Scaled dot-product attention scores are softmax-normalized and used to compute weighted value combinations. The output is reshaped, flattened, and passed through a three-layer FC block with ReLU activation after the first two layers, producing the final channel activity prediction.
	
	To quantitatively evaluate the activity level of each EEG channel, the Channel Activity Coefficient (CAC) is proposed. This metric incorporates three components: the classification accuracy and precision of the CNN model, as well as the Shannon entropy of its predictions. A summary of these components is provided in Table \ref{tab:1}. The CAC formulation is grounded in theoretical considerations and comprises three key metrics: the Relative Discriminatory Quotient (RDQ), the Accuracy-Chance Ratio (ACR), and Shannon entropy. The RDQ captures the relative discriminative capability of a channel by comparing its performance to the best and worst observed cases, while ACR measures deviations from chance-level accuracy, penalising predictions that approximate random guessing. A \textit{base-2} logarithm is applied to ACR to map its values onto an interpretable scale, where positive values indicate improvement over chance and negative values indicate sub-chance performance. Shannon entropy, used in the form $(1-H_i)$, reflects the confidence of the model's predictions--with lower entropy indicating higher certainty. These three quantities are then combined as shown in equation \ref{eq:1} to yield the CAC.
	
	\begin{table}[!tb]
		\centering
		\caption{Description of the components of the CAC metric.}
		\label{tab:1}
		\footnotesize
		\renewcommand{\arraystretch}{1.0}
		\setlength{\tabcolsep}{3pt}
		\begin{tabularx}{\textwidth}{|X|p{0.35\textwidth}|}
			\hline
			\textbf{Description} & \textbf{Formula} \\
			\hline
			\textbf{RDQ}: Measures how well a channel performs relative to the best and worst within the population. & 
			$(A_i - A_{\min}) / (A_{\max} - A_{\min} + \epsilon)$ \\
			
			\textbf{ACR}: Highlights deviation from chance-level accuracy (0.5); the base-2 logarithm normalizes the value scale. & 
			$\log_2((A_i + \epsilon) / (0.5 + \epsilon))$ \\
			
			\textbf{Entropy (H)}: Captures prediction confidence using Shannon entropy; lower values indicate more confident predictions. & 
			$-\sum_{c=1}^{2} p_c \log_2(p_c + \epsilon)$ \\
			\hline
		\end{tabularx}
	\end{table}
	
	\begin{equation}
		\text{CAC}_i = \text{RDQ}_i \cdot \text{ACR}_i \cdot \left( 1 - {H_i} \right)
		\label{eq:1}
	\end{equation}
	
	Where, \textit{i} indicates channel-\textit{i}. This metric is then computed for all channels in the dataset and the top-K channels are selected. In this work, the number of selected channels \textbf{K} was set to 16, as this configuration provides a practical balance between signal utility and system simplicity. Sixteen-channel EEG caps are commonly used in BCI research and support key electrode layouts (e.g., from the 10–20 and 10–10 systems) that are sufficient for typical BCI tasks such as motor imagery and cognitive workload estimation.
	
	\section{Experiments and Results}
	\label{sec:3}
	
	The following set of experiments was conducted to evaluate the robustness and effectiveness of the proposed method:
	
	\begin{itemize}
		\item The base model was trained to discriminate between active and resting state EEG data.
		\item The CAC metric was computed for each individual channel using the predictions of the trained model.
		\item The top 16 channels were selected based on the CAC scores, and their performance was evaluated using standard benchmark classifiers.
		\item The performance obtained using the proposed ACCS method was compared against other established channel selection algorithms.
	\end{itemize}
	
	To conduct these experiments, imagined speech EEG was selected as the primary domain. This choice is motivated by the fact that imagined speech EEG is particularly sensitive to cross-channel attenuation, and significant research efforts have been dedicated to addressing this challenge. In this study, the KARAOne dataset \cite{zhao2015classifying}—an open-source dataset for imagined speech EEG—was used, and its details are provided in Table~\ref{tab:2}. Training and evaluation are conducted on a per-subject basis, with each subject's EEG data labeled as either active or inactive, corresponding to imagined speech and idle conditions, respectively. To compute the CAC metric, each EEG channel is processed independently by training the model using only that specific channel's data across all available trials. In other words, the model is trained to distinguish active from resting states using univariate time series corresponding to each channel in isolation, allowing the direct comparison of the discriminative power of individual channels. The dataset is partitioned into training, validation, and test sets using the \texttt{train\_test\_split} function from the \texttt{scikit-learn} library, in a 70:15:15 ratio. A batch size of 32 is used, and the training is run for 50 epochs. The model is optimized using binary cross-entropy loss and the Adam optimizer with an initial learning rate of 6$\times$10$^{-3}$, which is dynamically reduced by a factor of 0.1 using a ReduceLROnPlateau scheduler based on validation loss.
	
	\begin{table}[!tb]
		\caption{KARAOne dataset description \cite{zhao2015classifying}}
		\label{tab:2}
		\renewcommand{\arraystretch}{1.0}
		\begin{tabularx}{\textwidth}{|p{0.3\textwidth}|X|}
			\hline
			\textbf{Parameter} & \textbf{Description} \\
			\hline
			EEG system & 64-channel Neuroscan Quick-cap \\
			Electrode placement & International 10-20 system \\
			Acquisition & SynAmps RT amplifier with 1 kHz sampling frequency \\
			Number of classes & 11 classes (4 words, 7 phonemes), 12 trials per class \\
			Number of participants & 13 (1 session each) \\
			Duration recorded & 5 s each for active and rest state \\
			\hline
		\end{tabularx}
	\end{table}
	
	\begin{figure}[!tb]
		\centering
		\includegraphics[width=\linewidth]{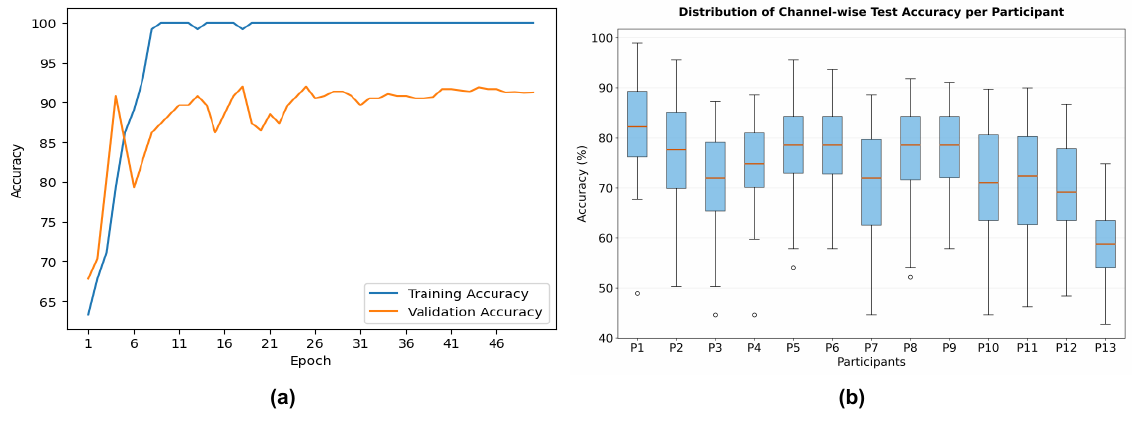}
		\caption{Base model performance: (a) training and validation accuracy over epochs, and (b) test accuracy distribution across channels for each participant.}
		\label{fig:2}
	\end{figure}
	
	\textbf{\textit{Experiment 1 --}} Figure~\ref{fig:2}(a) shows the training and validation performance of the model. Training accuracy saturates near 100\% within the first 10 epochs, while validation accuracy stabilizes above 90\%, indicating that the model generalizes well to unseen data. Figure~\ref{fig:2}(b) presents the distribution of test accuracies achieved using individual channels across participants. Although some inter-subject variability is observed, the majority of channels yield test accuracies well above chance, with several participants showing medians above 80\%. These results demonstrate that the model captures discriminative information from individual EEG channels and is suitable for evaluating channel importance in the ACCS framework.
	
	\begin{figure}[!tb]
		\centering
		\includegraphics[width=\textwidth]{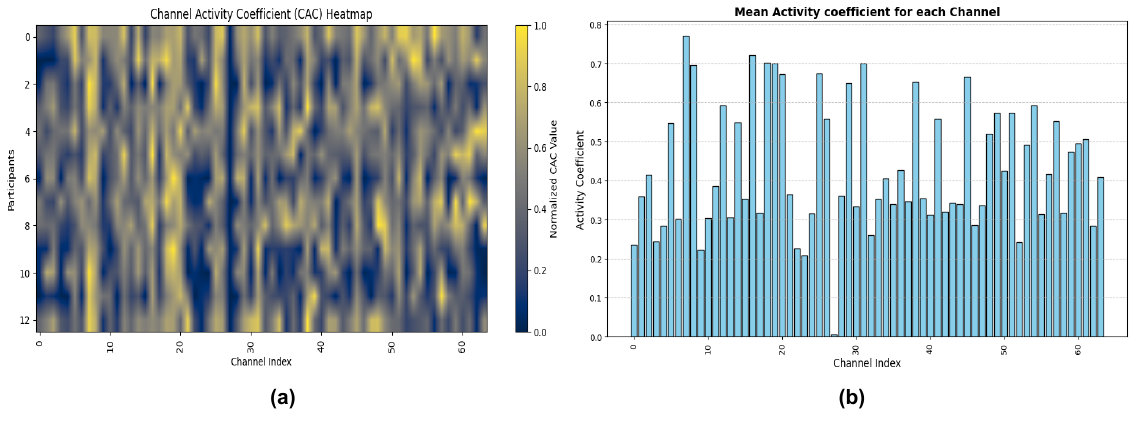}
		\caption{Visualization of the CAC metric across channels: (a) heatmap of CAC scores computed per participant, and (b) average CAC value for each channel, aggregated across participants.}
		
		\label{fig:3}
	\end{figure}
	
	\textbf{\textit{Experiment 2 --}} The Channel Activity Coefficient (CAC), introduced in Section~\ref{sec:2}, quantifies the discriminative power of each EEG channel. Figure~\ref{fig:3}(a) shows a heatmap of normalized CAC scores across all participants and channels. While some variability exists, a consistent pattern emerges where certain channels exhibit higher activity levels across subjects. Figure~\ref{fig:3}(b) presents the mean CAC value per channel, highlighting a clear disparity: several channels consistently show low activity, indicating limited contribution beyond baseline signals, while others stand out with notably higher scores. Based on this distribution, the top 16 channels with the highest average CAC values are selected for further evaluation.
	
	\begin{table}[!tb]
		\centering
		\caption{Accuracy improvement (\%) with channel selection over baseline (64 ch.).}
		\label{tab:3}
		\renewcommand{\arraystretch}{1.0}
		\begin{tabularx}{\textwidth}{|p{0.07\textwidth}|p{0.35\textwidth}*{6}{|>{\centering\arraybackslash}X}|}
			\hline
			\textbf{Ref.} & \textbf{Model} & \textbf{B1} & \textbf{B2} & \textbf{B3} & \textbf{B4} & \textbf{B5} & \textbf{MC} \\
			\hline
			\cite{saha2019deep} & CrosCov Hybrid ConvLSTM & 5.13 & 3.82 & 12.99 & 5.48 & 7.68 & 33.40 \\
			\cite{mini2021eeg} & Cepstral ANN & 4.02 & 8.47 & 12.81 & 10.28 & 1.87 & 26.60 \\
			\cite{rusnac2022cnn} & Spectral Windowed CNN & 6.68 & 7.67 & 9.41 & 7.81 & 4.29 & 27.81 \\
			\cite{rousis2024combining} & EEGNet+SPDNet & 10.31 & 10.45 & 8.94 & 13.61 & 3.18 & 39.52 \\
			\cite{Carvalho.VR.2024} & DDA+SVM & 9.06 & 7.79 & 10.07 & 7.96 & 4.32 & 34.97 \\
			\hline
		\end{tabularx}
	\end{table}
	
	% \textbf{\textit{Experiment 3 --}} To assess the impact of the proposed ACCS method, five state-of-the-art models were evaluated on the KARAOne dataset for imagined speech classifications which included 5 binary classification - B1(), B2(), B3(), B4(), B5(), and a 11-label multiclass classification(MC) as shown in Table~\ref{tab:3}. With ACCS, the number of channels was reduced to 16, while classification performance consistently improved across all models and tasks. For instance, DDA+SVM \cite{Carvalho.VR.2024} improved from 29.65\% to 40.02\% on MC, Cepstral ANN \cite{mini2021eeg} improved from 88.7\% to 93.6\%, and EEGNet+SPDNet \cite{rousis2024combining} improved from 84.9\% to 93.8\% on average in Binary classification. The CrosCov Hybrid ConvLSTM model \cite{saha2019deep} achieved the highest observed accuracy of 96.6\% on the B5 task. These consistent gains highlight ACCS’s ability to enhance performance by retaining only the most informative EEG channels.
    
	\textbf{\textit{Experiment 3 --}}To evaluate the effectiveness of the proposed ACCS method, five state-of-the-art models were tested on the KARAOne dataset for five binary (B1–B5) and one multiclass (MC) imagined speech classification tasks. As shown in Table~\ref{tab:3}, ACCS reduced the number of EEG channels to 16 while consistently improving accuracy across all models. Notably, EEGNet+SPDNet achieved a 39.52\% improvement on the MC task, while DDA+SVM achieved a 34.97\% gain. For binary tasks, Cepstral ANN \cite{mini2021eeg} improved from 88.7\% to 93.6\%, and EEGNet+SPDNet \cite{rousis2024combining} improved from 84.9\% to 93.8\%. On average, the highest improvement (32.46\%) occurred in the MC task, likely due to its complexity and greater benefit from noise reduction. The lowest gain was seen in B5 (4.26\%), though still positive.These consistent gains highlight ACCS’s ability to enhance performance by retaining only the most informative EEG channels. 
	
	\begin{table}[!tb]
		\centering
		\caption{Comparison of the performance improvement (\%) over the baseline.}
		\label{tab:4}
		\renewcommand{\arraystretch}{1.0}
		\begin{tabularx}{\textwidth}{|p{0.07\textwidth}|p{0.15\textwidth}|>{\centering\arraybackslash}p{0.08\textwidth}*{6}{|>{\centering\arraybackslash}X}|}
			\hline
			\textbf{Ref.} & \textbf{Method} & \textbf{TI} & \textbf{B1} & \textbf{B2} & \textbf{B3} & \textbf{B4} & \textbf{B5} & \textbf{MC} \\
			\hline
			\cite{1615389} & PCA & Y & 1.74 & 4.24 & 6.81 & 5.69 & 3.63 & 12.85 \\
			\cite{wang2019channel} & MI & N & 7.81 & 6.43 & 8.78 & 7.36 & 3.59 & 21.52 \\
			\cite{wang2020eca} & ECA-Net & N & 9.58 & 6.20 & 11.45 & 6.85 & 4.63 & 33.96 \\
			\cite{moctezuma2020eeg} & NSGA-II & N & 8.50 & 5.93 & 7.78 & 8.47 & 4.09 & 37.17 \\
			\cite{yu2021cross} & XCDC & N & 8.76 & 7.48 & 7.41 & 6.50 & 3.88 & 24.69 \\
			\cite{tang2022eeg} & SBFS & N & 7.44 & 6.22 & 9.25 & 8.47 & 2.20 & 29.85 \\
			\hline
			--- & \textbf{ACCS} & \textbf{Y} & \textbf{9.06} & \textbf{7.79} & \textbf{10.07} & \textbf{7.96} & \textbf{4.32} & \textbf{34.97} \\
			\hline
		\end{tabularx}
	\end{table}
	
	\textbf{\textit{Experiment 4 --}} The effectiveness of ACCS was further evaluated by comparing it with several established channel selection methods, as summarized in Table~\ref{tab:4}. All methods were tested using the DDA+SVM \cite{Carvalho.VR.2024} classifier, and results are reported as relative improvements over the 64-channel baseline. The proposed method achieved achieved one of the highest average improvements (34.97\%) and outperformed all other methods in four out of six tasks, including B1 (9.06\%), B2 (7.79\%), B3 (10.07\%), and B5 (4.32\%). Although NSGA-II \cite{moctezuma2020eeg} achieved the best result on the multi-class task (40.67\%), ACCS remained competitive and demonstrated more consistent improvements overall. Unlike the other approaches, which are task-specific and depend on classifier feedback, ACCS performs task-agnostic selection without classifier involvement, allowing the same subset of channels to be reused across tasks, while significantly outperforming traditional task-independent methods like PCA.
	
	\section{Conclusion}
	This work presents a task-agnostic channel selection algorithm for reducing the computational demands of EEG-based imagined speech classification. The proposed method selects a fixed subset of 16 channels by quantifying the activity level of each channel using a novel metric. This enables consistent performance improvements across multiple models and tasks. In contrast to existing approaches that rely on feedback from specific classifiers, the method operates independently of task or model, allowing the same channel subset to be reused across different settings. The experimental results confirm its effectiveness and demonstrate competitive performance relative to state-of-the-art techniques. It is important to note, however, that this approach is tailored for BCI applications and may not be suitable for clinical EEG data, where idle-state signals can contain diagnostically meaningful information.
	
	\bibliographystyle{splncs04}
	
\end{document}